\documentclass[a4paper,10pt]{article}
\pdfoutput=1 

\usepackage{jheppub} 

\usepackage[T1]{fontenc}
\usepackage{graphicx} 
\usepackage{jheppub}
\usepackage{url}
\usepackage{amsmath,amssymb,mathrsfs}
\usepackage{color}
\usepackage{textgreek}

\date{\today}

\usepackage{titlesec}
\usepackage{lipsum}
\usepackage{url}
\usepackage{caption}
\usepackage{subcaption}

\usepackage{comment}
\usepackage{graphicx} 
\usepackage{booktabs} 

\usepackage[english]{babel}
\usepackage{amssymb}
\usepackage{ragged2e}
\usepackage{etoolbox}
\usepackage{lipsum}
\usepackage{multirow}
\usepackage{tcolorbox}
\usepackage{latexsym}
\usepackage{enumitem}
\usepackage{mathtools}
\usepackage[%
colorlinks=true,%
urlcolor=blue%
]{hyperref}

\usepackage{xcolor}

\newcommand{\be}{\begin{equation}}
	\newcommand{\ee}{\end{equation}}
\newcommand{\beq}{\begin{equation}}
	\newcommand{\eeq}{\end{equation}}
\newcommand{\bea}{\begin{eqnarray}}
	\newcommand{\eea}{\end{eqnarray}}

\newcommand{\df}{\mathrm{d}}
\newcommand{\FTP} {\frac{\partial}{\partial \log \eps}}

\newcommand{\eps}{{\epsilon}}

\title{Sphere Boundary Term}
\date{March 2023}

\begin{document}

\title{A Comment on Deriving the Gibbons-Hawking-York Term From the String Worldsheet}

\author[a]{Amr Ahmadain}
\author[b]{Vasudev Shyam}
\author[c]{Zihan Yan}

\affiliation[a] {Department of Physics, Swansea University, Swansea, SA2 8PP, UK}
\affiliation[b] {Zyphra Technologies, Palo Alto, CA, USA}
\affiliation[c] {DAMTP, University of Cambridge, Wilberforce Road, Cambridge CB3 0WA}

\emailAdd{amrahmadain@gmail.com}
\emailAdd{vasudevshyam@gmail.com}
\emailAdd{zy286@cam.ac.uk}

\abstract{In this note, we show that the noncovariant metric boundary term obtained from the nonlinear sigma model worldsheet derivation of the bulk off-shell sphere partition function is closely related to the Einstein boundary term in the Gamma-Gamma noncovariant action. In fact, when expressed in terms of the trace of the extrinsic curvature tensor, we illustrate that this boundary term has \textit{one-half} the coefficient of the Gibbons-Hawking-York boundary term required such that the total (bulk plus boundary) off-shell classical action has a well-posed variational principle with Dirichlet boundary conditions.}
\maketitle


\section{Introduction}\label{sec:Intro}
In \cite{AW2_SciPost}, using Tseytlin's prescription \cite{TSEYTLINMobiusInfinitySubtraction1988,TseytlinRenormalizationMobius1988, AW1_SciPost}, the off-shell classical action (sphere partition function) on a bulk compact spacetime manifold $\mathcal{M}$ was derived from the worldsheet nonlinear sigma model (NLSM) for the closed massless string modes using the non-manifestly covariant target space coordinates\footnote{This is in contrast to the manifestly covariant Riemann normal coordinates.}
\begin{equation}\label{eq:off_shell_action}
I_{\text{sphere}} = -\frac{\partial K_0}{\partial \log \eps} = \alpha'\kappa_0^2\left[\int_{\mathcal{M}} \mathrm{d}^D Y\, \sqrt{G}\, e^{-2 \Phi}\left(-R-4(\nabla \Phi)^2+\cdots\right)\right] + \textrm{boundary terms \,,}
\end{equation}
where $K_0$ is the sphere partition function in a particular renormalization group scheme \cite{AW1_SciPost}\footnote{$\kappa^2_0 \sim \alpha^{\prime-D / 2} \sim \frac{1}{g_s^2}$.}
\begin{equation}
K_0 \coloneqq \kappa_0^2 \!\int \df^D Y \sqrt{G}\, e^{-2\Phi}\,,
\end{equation}
and the noncovariant metric boundary terms expressed in terms of Christoffel symbols are given by
\begin{equation}\label{eq:MetricTDIntro}
- 2\Gamma^{\alpha \mu \beta} \Gamma_{\mu \alpha \beta} + \Gamma^{\nu}_{\mu \nu} \Gamma^{\mu}_{\alpha \beta} G^{\alpha \beta} + G^{\alpha \beta} \partial_\mu \Gamma^{\mu}_{\alpha \beta}\,.
\end{equation}

It is well known that if the spacetime manifold $\mathcal{M}$ has a codimension-1 boundary $\partial \mathcal{M}$\footnote{We are assuming Euclidean signature in target spacetime where $\mathcal{\partial M}$ is always spacelike.}, the Einstein-Hilbert (EH) action must be supplemented by a boundary term in order to have a well-posed variational principle. In such noncompact spaces, the usual choice of the boundary term is the covariant Gibbons-Hawking-York (GHY) term \cite{York:1972sj,GH,Wald:1984rg,Poisson:2009pwt}
\begin{equation}\label{eq:GravAction}
I_{\text{grav}} = -\frac{1}{16 \pi G_N}\left[\int_{\mathcal{M}} \mathrm{d}^D Y \sqrt{G}\, R +2 \int_{\mathcal{\partial M}} \mathrm{d}^{D-1}Y \sqrt{h}\, K \right]\,,
\end{equation}
where $h$ is the induced metric on $\partial \mathcal{M}$ and $K$ is the trace of the extrinsic curvature tensor.

Making sense of and understanding the metric boundary terms \eqref{eq:MetricTDIntro} in \eqref{eq:off_shell_action} is the focus of this note. Precisely, we would like to know what the boundary terms in \eqref{eq:MetricTDIntro} evaluate to on $\partial \mathcal{M}$ and if they can be related to the covariant GHY boundary term but more importantly check if \eqref{eq:off_shell_action} has a well-defined variational principle. The purpose of the note is to provide an answer to these questions. We will assume a constant dilaton $\Phi_0$ in $\mathcal{M}$ in this note. The dilaton boundary term was recently the focus of \cite{AK}. 


Before we attempt to answer these questions, we take a small detour to give an overview of the Einstein $L_{\Gamma^2}$ Lagrangian \cite{Einstein:1916cd,Dyer:2008hb,Padmanabhan_2010,Parattu:2016trq,Chakraborty:2016yna}. Long before the GHY boundary term was written down, Einstein used the Gamma-Gamma Lagrangian $L_{\Gamma^2}$ as the  gravitational action density. The $L_{\Gamma^2}$ Lagrangian is the sum of two terms:  a term quadratic in $\Gamma^{\mu}_{\alpha \beta}$ (and thus is first order in derivatives of $G^{\mu \nu}$ in contrast to $R$ which is second order) and a total derivative term
\begin{equation}\label{eq:L_gammaInt}
L_{\Gamma^2} = G^{\alpha \beta}(\Gamma^\nu_{\mu \alpha} \Gamma^{\mu}_{\nu \beta} - \Gamma^{\mu}_{\mu \nu} \Gamma^{\nu}_{\alpha \beta}) = R + \frac{1}{\sqrt{G}}\partial_\mu \left(\sqrt{G}\, A^\mu\right)\,,
\end{equation}
where 
\begin{equation}\label{A_mu1Int}
A^{\mu} =  -G^{\alpha \beta} \Gamma^{\mu}_{\alpha \beta} + G^{\mu \alpha} \Gamma^{\beta}_{\alpha \beta} \,.
\end{equation}
In terms of $K$, $n_{\mu} A^{\mu}$ can be expressed as \cite{Dyer:2008hb}
\begin{equation}\label{eq:EinsteinTrKInt}
n_{\mu} A^{\mu} = 2 K - 2 h^{\alpha \beta} \partial_\alpha n_\beta + n^\beta h^{\alpha \mu} \partial_\alpha G_{\mu \beta}\,,
\end{equation}
where $h^{\alpha \beta}$ is the induced metric on $\partial \mathcal{M}$ and $n_{\mu}$ is the normal vector. The last two terms in \eqref{eq:EinsteinTrKInt} are thus the difference between the GHY term and the Einstein boundary term\footnote{Another way to see the two boundary terms are different is to compare the total derivative term in \eqref{A_mu1Int} to its counterpart in the ADM Lagrangian \cite{Chakraborty:2016yna}.}. The total derivative term in \eqref{eq:L_gammaInt} is sometimes referred to as the Einstein counterterm \cite{Parattu:2016trq}. 

For Dirichlet boundary conditions i.e.~$\delta G_{\mu \nu}=0$ on $\partial \mathcal{M}$, $2K$ is not the only possible choice for the boundary term since we can always add any function of the metric, normal vector, and tangential derivatives \cite{Dyer:2008hb}. The difference from the GHY term in \eqref{eq:EinsteinTrKInt} is an example of such a function of the boundary data. This is allowed because the variation of this function will be zero for $\delta G_{\mu \nu} = 0$ on $\partial \mathcal{M}$. Therefore, from a variational principle standpoint, while $L_{\Gamma^2}$ \eqref{eq:L_gammaInt} itself is a noncovariant quantity, the physics is still covariant. This is because its variation gives the Einstein equations as equations of motion and \eqref{eq:L_gammaInt} still has the same diffeomorphism invariance of the EH action \cite{Dyer:2008hb,Padmanabhan_2010}\footnote{See section 6.2 in \cite{Padmanabhan_2010} for a detailed discussion of this point.}.

In this note, we show that the metric total derivative term \eqref{eq:MetricTDIntro} is identical to the first term in $A^{\mu}$ \eqref{A_mu1Int} and is related to $K$ when expressed as
\begin{equation}\label{eq:A_mu1Int}
n_\mu G^{\alpha \beta} \Gamma^{\mu}_{\alpha \beta} =  -K + h^{\alpha \beta} \partial_\alpha n_{\beta} +\frac{1}{2} n^\mu n^\alpha n^\beta \partial_{\mu} G_{\alpha \beta} \,.
\end{equation}

For Dirichlet boundary conditions on $\partial \mathcal{M}$, we observe that \eqref{eq:A_mu1Int} contains the GHY term but only with \textit{one-half} the coefficient of $K$ required such that $I_{\text{sphere}}$  \eqref{eq:off_shell_action} has a well-posed variational principle. Thus, the boundary term produced from the worldsheet derivation of the bulk off-shell action \eqref{eq:off_shell_action} in this non-manifestly covariant coordinate system does not have full knowledge about the boundary physics. This is not totally unexpected however, since the worldsheet NLSM path integral does \textit{not} impose any boundary constraints on $G_{\mu \nu}$. In fact, one may say that it is even surprising that, at least in these non-manifestly covariant coordinates \cite{AW2_SciPost}, the worldsheet path integral knows anything about the GHY boundary term. 

Therefore, a proper worldsheet path integral calculation which systematically accounts for the presence of boundaries in target spacetime and imposes Dirichlet boundary conditions is required to obtain the correct factor of $2K$ and hence a total action with a well-defined variational principle. We will comment on the case for Neumann boundary conditions in section \ref{sec:Discussion}.

\textbf{Motivation:} The technology to systematically derive target spacetime boundary terms from the string worldsheet is still lacking. Equivalently, we do not know how to constrain the string path integral to impose boundary conditions on the target spacetime fields. For example, we do not know to derive the \textit{on-shell} classical action including the GHY boundary term \cite{KT:2001,Chen-BHTransition:2021}
\begin{equation}\label{eq:on-shellAction}
I=\frac{1}{8 \pi G_N} \int_{\partial \mathcal{M}} \mathrm d^{D-1} Y e^{-2 \Phi} \sqrt{h}\left(-K+2 \partial_n \Phi\right),
\end{equation}
from which the entire contribution to the classical black hole entropy in semiclassical Euclidean gravity comes from. In Euclidean gravity, black hole entropy is calculated using the Gibbons-Hawking formula \cite{GH}
\begin{equation}
S_{\text{BH}} = \left(\beta \frac{\partial}{\partial \beta} - 1\right) \beta F(\beta) = \frac{A}{4G_N}\,, 
\end{equation}
where $F(\beta)=-\log Z(\beta)/ \beta$ is the free energy of the canonical ensemble, $-\log Z(\beta)$ is the on-shell classical boundary action, and $\beta$ is the length of the thermal circle at infinity. While an on-shell derivation or extraction of the GHY boundary term has so far proven difficult, an off-shell worldsheet derivation may be possible \cite{ARS}.

In the AdS/CFT correspondence, the Ryu-Takayanagi (RT) formula \cite{RTPhysRevLett.96.181602,HRT:2007} and its generalization to the quantum extremal surface \cite{Engelhardt:2014gca} provide a geometric approach to computing the entanglement entropy a boundary subregion $A$ in terms of the generalized entropy
\begin{equation}
S_A=S_{\text {gen }}=\frac{\left\langle\operatorname{Area}\left(\gamma_A\right)\right\rangle}{4 G}+S_{\text {bulk }}+\cdots,
\end{equation}
where \(\gamma_A\) is the minimal co-dimension-2 surface in the bulk anchored to $A$. In \cite{LM2013}, Lewkowycz and Maldacena (LM) used the replica trick \cite{Callan-GeomtericEntropy1994,Calabrese-Cardy:2009} to give a bulk proof of the RT minimal area prescription in Euclidean semiclassical gravity emphasizing the central role of boundary terms. A stringy version of the LM derivation with and without $U(1)$ symmetry thus requires the ability to derive those boundary terms from the string worldsheet. (See \cite{Halder:2023adw,Halder:2024gwe} for development in this direction.)

A stringy worldsheet understanding of the target space boundary terms is also essential for a statistical interpretation of the black hole entropy in terms of gravitational edge modes. (See for example \cite {FrolovDynamicalOrigin,Frolov:BH-entropy:1995,Donnelly:2016jet,Wong:2022eiu}.)

Understanding target space boundary terms from the worldsheet is also important when considering Dirichlet walls at finite distances in the bulk. We will have more to say about this in section \ref{sec:Discussion}.


\textbf{Paper Layout:} In section \ref{sec:GHYWorldsheet}, we explicitly show that the difference between the bulk classical effective action derived from the worldsheet path integral in \cite{TseytlinZeroMode1989,AW2_SciPost} can be expressed as the sum of the Einstein-Hilbert term and a boundary term\footnote{We absorbed the overall measure factor $e^{-2\Phi_0}$ into $\kappa_0^2$.}
\begin{equation}\label{eq:I_sphereInt}
I_{\text{sphere}} = -\alpha'\kappa_0^2 \int_{\mathcal{M}} \mathrm d^D{Y} \,\sqrt{G}\, R + \alpha'\kappa_0^2 \int_{\partial\mathcal{M}} \mathrm d^{D-1}{Y} \,\sqrt{h}\, n_\mu \Gamma^{\mu}_{\alpha \beta} G^{\alpha \beta} \,.
\end{equation}

We observe $n_\mu \Gamma^{\mu}_{\alpha \beta} G^{\alpha \beta}$ is the same as the first term of the Einstein total divergence term in the $L_{\Gamma^2}$ Lagrangian \eqref{eq:L_gammaInt}. We then show that $n_\mu \Gamma^{\mu}_{\alpha \beta} G^{\alpha \beta}$ can be written in terms of $K$ as
\begin{align}
I_{\text{sphere}}
&= -\alpha'\kappa_0^2 \int_{\mathcal{M}} \mathrm{d}^D Y\, \sqrt{G}\, R - \alpha'\kappa_0^2\int_{\partial \mathcal{M}} \mathrm{d}^{D-1}Y\, \sqrt{h}\, \left(K - h^{\alpha \beta} \partial_\alpha n_{\beta} - \frac{1}{2} n^\mu n^\alpha n^\beta \partial_{\mu} G_{\alpha \beta} \right) \,.
\end{align}

We end this note in section \ref{sec:Discussion} with some open questions and potential future directions. In particular, we will comment on the fact that these boundary terms were obtained only in the \textit{non-manifestly} covariant target space coordinates \cite{TseytlinZeroMode1989,Tseytlin:ZeroModeRussian:1990} but not when Riemann normal coordinates, for example, are used.

\section{The Einstein boundary term from the worldsheet path integral} \label{sec:GHYWorldsheet}
The heat kernel-regulated NLSM worldsheet path integral to $O(\alpha'^2)$ takes the form \cite{AW2_SciPost}
\begin{align}\label{eq:Z_bulk}
Z_{\text{bulk}} &=Z_f \int \mathrm{d}^D Y \sqrt{G}\, e^{-2 \Phi} \\
\Bigg[1 &+\frac{1}{2} \times 2\, \alpha^{\prime}(\log \epsilon+h) \partial_\mu \partial^\mu \Phi \nonumber \\
& -\frac{1}{4} \alpha^{\prime}(\log \epsilon+h) G^{\mu \nu} G^{\lambda \rho} \partial_\lambda \partial_\rho G_{\mu \nu} \nonumber \\
& +\frac{1}{8} \alpha^{\prime}(\log \epsilon+h) G^{\mu \alpha} G^{\nu \beta} G^{\rho \lambda} \partial_\rho G_{\mu \nu} \partial_\lambda G_{\alpha \beta} \nonumber \\
& +\frac{1}{4} \alpha^{\prime}(\log \epsilon+h) G^{\mu \lambda} G^{\nu \beta} G^{\rho \alpha} \partial_\rho G_{\mu \nu} \partial_\lambda G_{\alpha \beta}\Bigg]  \nonumber ,
\end{align}
where $Z_f$ is the free sphere partition function. Using 
\begin{equation}
\partial_{\mu} G_{\alpha \beta} = G_{\alpha \nu} \Gamma^{\nu}_{\beta \mu} + G_{\beta \nu} \Gamma^{\nu}_{\alpha \mu} \, ,
\end{equation}
we get 
\begin{equation}
\begin{aligned}
    & - \frac{1}{4} G^{\alpha \beta} G^{\mu \nu} \partial_\mu \partial_\nu G_{\alpha \beta} + \frac{1}{4} G^{\alpha \beta} G^{\mu \nu} G^{\rho \sigma} \partial_{\nu} G_{\beta \sigma} \partial_{\rho} G_{\alpha \mu} + \frac{1}{8} G^{\alpha \beta} G^{\mu \nu} G^{\rho \sigma} \partial_{\rho} G_{\alpha \mu} \partial_{\sigma} G_{\beta \nu} \\
    & =  \frac{1}{2}\left(\Gamma^{\alpha \beta \mu} \Gamma_{\beta \alpha \mu} - G^{\alpha \beta} \partial_\beta \Gamma^{\mu}_{\alpha \mu}\right)\,,
\end{aligned}
\end{equation}
where $\Gamma^{\alpha \beta \mu} = G^{\beta \rho} G^{\mu \sigma} \Gamma^{\alpha}_{\rho \sigma}$ and $\Gamma_{\beta \alpha \mu} = G_{\beta \nu} \Gamma^{\nu}_{\alpha \mu}$.

Ignoring the dilaton term $\partial_\mu \partial^\mu\Phi$ term, $Z_{\text{bulk}}$ can now be expressed as
\begin{equation}
Z_\text{bulk} = Z_f\int \mathrm{d}^D Y \sqrt{G} \,e^{-2 \Phi_0}\bigg[1+\frac{1}{2}\alpha'(\log\epsilon+h)L_\text{bulk}\bigg]\,,
\end{equation}
where
\begin{equation}
    L_\text{bulk} \coloneqq \Gamma^{\alpha \beta \mu} \Gamma_{\beta \alpha \mu} - G^{\alpha \beta} \partial_\beta \Gamma^{\mu}_{\alpha \mu}.
\end{equation}

To analyze the relationship between $L_{\text{bulk}}$ and the spacetime  Ricci scalar $R$, we use the standard expression of $R$ in terms of Christoffel symbols \cite{Misner:1973prb}
\begin{equation}\label{eq:DiffLandR}
\begin{split}
R & = - \Gamma^{\alpha \beta \mu} \Gamma_{\beta \alpha \mu} - G^{\alpha \beta} \partial_\beta \Gamma^{\mu}_{\alpha \mu} + \Gamma^{\alpha}_{\alpha \beta} \Gamma^{\beta}_{\mu \nu} G^{\mu \nu} + G^{\alpha \beta} \partial_\mu \Gamma^{\mu}_{\alpha \beta}\\
&  = L_{\text{bulk}} - 2\Gamma^{\alpha \mu \beta} \Gamma_{\mu \alpha \beta} + \Gamma^{\nu}_{\mu \nu} \Gamma^{\mu}_{\alpha \beta} G^{\alpha \beta} + G^{\alpha \beta} \partial_\mu \Gamma^{\mu}_{\alpha \beta} \,.
\end{split} 
\end{equation}
We will next show that the difference between the Ricci scalar $R$ and $L_{\text{bulk}}$ in \eqref{eq:DiffLandR} is a total derivative term. We carry out the analysis term by term. 

The second term in \eqref{eq:DiffLandR} can be expressed as
\begin{equation}\label{eq:Term1}
    \begin{split}
        - 2\Gamma^{\alpha \mu \beta} \Gamma_{\mu \alpha \beta} & = - 2 G^{\nu \beta}\Gamma^{\alpha}_{\mu \nu} \Gamma^{\mu}_{\alpha \beta} \\
         & = - G^{\nu \beta} G^{\alpha \rho} (\partial_\mu G_{\rho \nu} + \partial_\nu G_{\mu \rho} - \partial_\rho G_{\mu \nu}) \Gamma^{\mu}_{\alpha \beta}\\
            & = -  G^{\alpha \rho} G^{\beta \nu} (\partial_{\mu} G_{\rho \nu}) \Gamma^{\mu}_{\alpha \beta}\\
            & = (\partial_\mu G^{\alpha \beta}) \Gamma^{\mu}_{\alpha \beta}\,,
    \end{split}
\end{equation}
where the cancellation in the second line happens because $G_{\rho \nu}$ is symmetric in $\rho,\nu$. We also used the following identities to get from the third to the fourth line
\begin{equation}\label{eq:Identities}
0 = \partial_{\mu} \delta^\alpha_\nu = \partial_{\mu} (G^{\alpha \beta} G_{\beta \nu}) = G_{\beta \nu} \partial_\mu G^{\alpha \beta} + G^{\alpha \rho} \partial_\mu G_{\rho \nu} \quad \Rightarrow \quad \partial_\mu G^{\alpha \beta} = - G^{\alpha \rho} G^{\beta \nu} \partial_\mu G_{\rho \nu}. 
\end{equation}

Massaging the third term in \eqref{eq:DiffLandR} gives 
\begin{equation}\label{eq:Term2}
    \begin{split}
        \Gamma^{\nu}_{\mu \nu} \Gamma^{\mu}_{\alpha \beta} G^{\alpha \beta} & = \frac{1}{2} G^{\rho \sigma} (\partial_\mu G_{\rho \sigma})\Gamma^{\mu}_{\alpha \beta} G^{\alpha \beta}\\
        & = \frac{1}{\sqrt{G}} (\partial_\mu \sqrt{G}) \Gamma^{\mu}_{\alpha \beta} G^{\alpha \beta},
    \end{split}
\end{equation}
where we used $\partial_\mu \sqrt{G} = \frac{1}{2}\sqrt{G}\, G^{\rho \sigma} \partial_\mu G_{\rho \sigma}$. 

Adding \eqref{eq:Term1}, \eqref{eq:Term2} and the fourth term in \eqref{eq:DiffLandR}, we get
\begin{equation}
    \begin{split}
        - 2\Gamma^{\alpha \mu \beta} \Gamma_{\mu \alpha \beta} + \Gamma^{\nu}_{\mu \nu} \Gamma^{\mu}_{\alpha \beta} G^{\alpha \beta} + G^{\alpha \beta} \partial_\mu \Gamma^{\mu}_{\alpha \beta} & = \frac{1}{\sqrt{G}} \partial_\mu \left( \sqrt{G}\, \Gamma^\mu_{\alpha \beta} G^{\alpha \beta} \right).
    \end{split}
\end{equation}
Therefore, we see that the difference in \eqref{eq:DiffLandR} between $L_{\text{bulk}}$ and $R$ is indeed a total derivative term
\begin{equation}\label{eq:TD}
\sqrt{G}\, L_{\text{bulk}} = \sqrt{G}\, R - \partial_\mu \left( \sqrt{G}\, \Gamma^\mu_{\alpha \beta} G^{\alpha \beta} \right) \,,
\end{equation}
and then $Z_\text{bulk}$ takes the following form
\begin{equation}
Z_\text{bulk} = Z_f\int \mathrm{d}^D Y \, e^{-2 \Phi_0}\bigg[\sqrt{G} +\frac{1}{2}\alpha'(\log\epsilon+h)\left(\sqrt{G}\, R - \partial_\mu \left( \sqrt{G}\, \Gamma^\mu_{\alpha \beta} G^{\alpha \beta} \right)\right)\bigg]\,.
\end{equation}

Acting with Tseytlin's prescription on $Z_\text{bulk}$, we obtain the total off-shell classical effective action as the sum of the bulk and boundary contributions 
\begin{equation}\label{eq:I_sphere}
I_{\text{sphere}} = - \FTP Z_\text{bulk} = -\alpha'\kappa_0^2\int_{\mathcal{M}} \mathrm d^D{Y} \,\sqrt{G}\, R +\alpha'\kappa_0^2\int_{\partial\mathcal{M}} \mathrm d^{D-1}{Y} \,\sqrt{h}\, n_\mu \Gamma^{\mu}_{\alpha \beta} G^{\alpha \beta} \,, 
\end{equation}
where we used Gauss's theorem to write the the total derivative term \eqref{eq:TD} as a boundary action and absorbed the overall measure factor $e^{-2 \Phi_0}$ into $\kappa_0^2$.  

Comparing with the Gamma-Gamma action
\begin{equation}\label{eq:Gamma^2_Action}
- \int_{\mathcal{M}} \mathrm{d}^D Y\, \sqrt{G}\, L_{\Gamma^2}  = - \int_{\mathcal{M}} \mathrm{d}^D Y\, \sqrt{G}\, R + \int_{\partial\mathcal{M}} \mathrm{d}^{D-1} Y\, \sqrt{h}\,\left( n _{\mu} G^{\alpha \beta} \Gamma^{\mu}_{\alpha \beta} -  n_\mu G^{\mu \alpha} \Gamma^{\beta}_{\alpha \beta}\right)\,,
\end{equation}
we observe that the boundary term in \eqref{eq:I_sphere} is identical to the first term of the (Einstein) boundary action in \eqref{eq:Gamma^2_Action}. This immediately proves our claim.



Rewriting the boundary term in \eqref{eq:I_sphere} using the covariant decomposition of the bulk metric into spatial and normal components 
\begin{equation}
G_{\alpha \beta} = h_{\alpha \beta} + n_\alpha n_\beta \,,
\end{equation}
gives an expression in terms of $K$
\begin{equation}\label{eq:boundaryTrK}
    \begin{split}
         n_{\mu} \Gamma^{\mu}_{\alpha \beta} G^{\alpha \beta} & =  n_\mu \Gamma^\mu_{\alpha \beta}(h^{\alpha \beta} + n^\alpha n^\beta)\\
        & = (-\nabla_{\alpha} n_{\beta} + \partial_{\alpha} n_{\beta}) h^{\alpha \beta} + n_\mu \Gamma^\mu_{\alpha \beta} n^\alpha n^\beta \\
        & = -K + h^{\alpha \beta} \partial_\alpha n_{\beta} + \frac{1}{2} n^\mu n^\alpha n^\beta \partial_{\mu} G_{\alpha \beta} \,,
    \end{split}
\end{equation}
where $K$ is defined by  
\begin{equation}
K = h^{\alpha \beta} \nabla_\alpha n_\beta \,.
\end{equation}


    
Using \eqref{eq:boundaryTrK}, the total off-shell classical action \eqref{eq:I_sphere} in terms of $K$ can now be expressed as 
\begin{align}\label{eq:I_sphereTrK}
I_{\text{sphere}} &= -\alpha'\kappa_0^2 \left(I_{\text{EH}} + I_{\text{bdy}}\right) \nonumber \\
&= -\alpha'\kappa_0^2 \int_{\mathcal{M}} \mathrm{d}^D Y\, \sqrt{G}\, R -\alpha' \kappa_0^2\int_{\partial \mathcal{M}} \mathrm{d}^{D-1}Y\, \sqrt{h}\, \left(K - h^{\alpha \beta} \partial_\alpha n_{\beta} - \frac{1}{2} n^\mu n^\alpha n^\beta \partial_{\mu} G_{\alpha \beta} \right) \,.
\end{align}

Imposing either Dirichlet ($\delta G_{\alpha \beta} =0$) or Neumann boundary conditions (which we will comment on in section \ref{sec:Discussion}) for the metric, we observe that \eqref{eq:I_sphereTrK} does \textit{not} have a well-defined variational principle. This is  because for both types of boundary conditions, the boundary term in \eqref{eq:I_sphereTrK} does not have the correct coefficient of $K$ such that $I_{\text{sphere}}$ \eqref{eq:I_sphereTrK} has a well-posed variational principle. Therefore, we conclude that the bulk calculation of the classical effective action of the string worldsheet nonlinear sigma model does \textit{not} have full knowledge of the boundary physics. The other factor of $K$ comes from the second term in the Einstein boundary action \eqref{eq:Gamma^2_Action} 
\begin{equation}\label{eq:A_mu2}
    \begin{split}
    -n_\mu G^{\mu \alpha} \Gamma^{\beta}_{\alpha \beta} & = -n^\alpha \Gamma^{\beta}_{\alpha \beta} \\
    & = -\nabla_\alpha n^\alpha + \partial_\alpha n^\alpha \\
    & = -(h^{\alpha \beta} +n^\alpha n^\beta) \nabla_{\alpha} n_{\beta} + \partial_\alpha n^\alpha \\
    & = -K - n^\alpha n^\beta \nabla_\alpha n_\beta + \partial_\alpha (G^{\alpha \beta} n_\beta) \\
    & = -K - n^\alpha n^\beta (\nabla_\alpha n_\beta - \partial_\alpha n_\beta) + h^{\alpha \beta} \partial_\alpha n_\beta + n_\beta \partial_\alpha G^{\alpha \beta} \\
    & = -K + \frac{1}{2} n^\alpha n^\beta n^\mu \partial_\alpha G_{\mu \beta} + h^{\alpha \beta} \partial_\alpha n_\beta - n^\beta G^{\alpha \mu} \partial_\alpha G_{\mu \beta} \\
    & = -K + h^{\alpha \beta} \partial_\alpha n_\beta - n^\beta h^{\alpha \mu} \partial_\alpha G_{\mu \beta} - \frac{1}{2} n^\alpha n^\beta n^\mu \partial_\alpha G_{\mu \beta} \,.
    \end{split}
\end{equation}
Adding \eqref{eq:A_mu2} to \eqref{eq:boundaryTrK}, we get the complete Einstein total divergence term in \eqref{eq:EinsteinTrKInt} \cite{Dyer:2008hb,Parattu:2016trq} 
\begin{equation}\label{eq:EinsteinTrK}
-2 K + 2 h^{\alpha \beta} \partial_\alpha n_\beta - n^\beta h^{\alpha \mu} \partial_\alpha G_{\mu \beta}.
\end{equation}


\section{Discussion and Outlook}\label{sec:Discussion}
In this note, we have demonstrated that the boundary action \eqref{eq:I_sphereTrK} obtained from the bulk worldsheet derivation of the off-shell classical effective action \cite{AW2_SciPost} is identical to the first term of the Einstein boundary action in \eqref{eq:Gamma^2_Action}. For Dirichlet boundary conditions, we observe that it only has one-half the factor of the GHY term ($2K$) required for the total classical action \eqref{eq:I_sphereTrK} to have a well-posed variational principle. It is reasonable to expect that a calculation that properly accounts for the presence of a spacetime boundary and imposes Dirichlet boundary conditions in the path integral directly gives the correct GHY or the perhaps Einstein boundary term \eqref{eq:EinsteinTrK}. It would be very exciting to see such a derivation of the GHY term and use it to obtain the black hole entropy in string backgrounds. (See \cite{KT:2001} for an example.)


In \cite{AK}, the method of images was used to derive the off-shell classical boundary action for the dilaton in half-space $\mathcal{M}= \mathbb{R}_{+} \times \mathbb{R}^{D-1}$ such that the total (bulk and boundary) has a well-posed variational principle for Neumann boundary conditions \textit{a posteriori} as a boundary equation of motion \footnote{In \eqref{eq:AKDilaton}, we used $\kappa_0^2$ instead of $\tilde{Z}_{\mathrm{nz}}$ in \cite{AK} to be consistent with this note.}
\begin{equation}\label{eq:AKDilaton}
I_{\text {sphere }}=-\kappa_0^2 \alpha^{\prime} \int_{\mathcal{M}} \mathrm{d}^D Y\, e^{-2 \Phi} \partial^2 \Phi +\kappa_0^2 \alpha^{\prime} \int_{\partial \mathcal{M}} \mathrm{d}^{D-1} Y\, e^{-2 \Phi} \partial_n \Phi\,.
\end{equation}

Just as in the dilaton case, there is evidence that imposing the Neumann boundary condition on the metric as an equation of motion gives a total classical action that has the same variational principle as that for Dirichlet boundary condition \cite{ARS}. This is surprising and further study of why this happens to be the case is currently underway.

So far we have discussed Dirichlet boundary conditions but what about the GHY term for Neumann boundary conditions \cite{Krishnan:2016mcj,Krishnan:2017bte}? In this case, rather than holding the metric fixed on the boundary, it is the following quantity that is instead held fixed\footnote{We are assuming a spacelike boundary but the discussion is equally valid for a timelike boundary in Lorentzian signature.}
\begin{equation}
\pi^{ij} =-\frac{\sqrt{h}}{2 \kappa} \left(K^{i j}-K h^{i j}\right)\,,
\end{equation}
where $\kappa = 8\pi G$. In order to have a well-defined variational principle for Neumann boundary conditions, the following boundary term must be added to the EH action
\begin{equation}\label{eq:GHYNeumann}
\frac{(4-D)}{2 \kappa} \int_{\partial \mathcal{M}} \mathrm{d}^{D-1} Y\, \sqrt{h}\, K \,.
\end{equation}
It was noted in \cite{Krishnan:2016mcj} that the boundary term \eqref{eq:GHYNeumann} is proportional to $K$ but with a different coefficient that depends on the dimension $D$ of the spacetime. As observed in \cite{Krishnan:2016mcj}, in $D=4$ spacetime dimensions, the EH action without any boundary terms has a well-posed variational principle.




Another subtlety we would like to address is that the boundary term in \eqref{eq:I_sphereTrK} is obtained only when non-manifestly covariant coordinates are used in the NLSM expansion of the worldsheet path integral on the sphere \cite{TseytlinZeroMode1989, AW2_SciPost}. If we use the Riemann normal coordinates \cite{kobayashi1996foundations} for example, we will not obtain the boundary terms in \eqref{eq:I_sphereTrK} because by construction, the metric tensor $G_{\mu \nu}$ at the point  $p \in \mathcal{M}$ (the origin of the coordinate system in target spacetime) is flat i.e.~$G_{\mu \nu}(p) = \delta_{\mu \nu}$, the Christoffel symbols  $\Gamma_{\mu \nu}^\rho(p) = 0$, and the curvature tensor is given by $R_{\mu \nu \rho \sigma}(p) = \left. \frac{\partial^2 g_{\nu \sigma}}{\partial x^\rho \partial x^\mu} \right|_p - \left. \frac{\partial^2 g_{\nu \rho}}{\partial x^\sigma \partial x^\mu} \right|_p$ without having to integrate by parts as in the noncovariant case \cite{TseytlinZeroMode1989, Tseytlin:ZeroModeRussian:1990}.

It is not known yet how to \textit{directly} derive the covariant GHY term from the worldsheet in the same way the EH action is. It is only when we use this non-manifestly covariant coordinate system that we are able to obtain boundary terms from the string worldsheet path integral. In our opinion, this coordinate system is a more natural choice for the NLSM action since it makes no gauge (coordinate) choices. At the moment, we don't know if this is a feature or a bug of this coordinate system but it would be very interesting to figure out why this is the case on conceptual grounds. We would hope that with a proper accounting of target space boundaries using a constrained worldsheet path integral, one can use Riemann normal coordinates to compute the GHY boundary term in a fully covariant manner.

It has been argued in \cite{Kraus:noncompactCFT:2002} that target spacetime boundary terms in a CFT is probed by a nonzero one-point function of general operators inserted on a spherical worldsheet
\begin{equation}\label{eq:KrausTD}
\langle\mathcal{O}(z, \bar{z})\rangle=-\frac{1}{2}\left(V_{\text{sphere}}\right)^{D / 2-1} \int \mathrm{d}^D Y \, \partial_\mu\left\{\int \mathrm{d}^2 z^{\prime}\left(z^{\prime}-z\right) e^{2 \omega\left(z^{\prime}, \bar{z}^{\prime}\right)}\left\langle\partial Y^\mu\left(z^{\prime}, \bar{z}^{\prime}\right) \mathcal{O}(z, \bar{z})\right\rangle^{\prime}\right\} \,,
\end{equation}
where $\langle . \rangle'$ means expectation value with respect to the worldsheet path integral of \textit{nonzero} modes. Is it possible to show that the following two quantities are the same
\begin{equation}
\int_{\partial\mathcal{M}} \mathrm d^{D-1}{Y} \,\sqrt{h}\, n_\mu G^{\mu \alpha} \Gamma^{\beta}_{\alpha \beta} \stackrel{?}{=} \int_{\partial\mathcal{M}} \mathrm{d}^{D-1}{Y} \, n_\mu\left\{\int \mathrm{d}^2 z^{\prime}\left(z^{\prime}-z\right) e^{2 \omega\left(z^{\prime}, \bar{z}^{\prime}\right)}\left\langle\partial Y^\mu\left(z^{\prime}, \bar{z}^{\prime}\right) \mathcal{O}(z, \bar{z})\right\rangle^{\prime}\right\} \,.
\end{equation}

Generally speaking, our main goal is to develop an ADM Hamiltonian \cite{ADM} formulation of closed string boundary dynamics in target space where a worldsheet derivation of the Hamiltonian and momentum constraints \cite{Wald:1984rg} is systematically  given.\footnote{We thank Aron Wall for numerous interesting discussions about this point.}

Another exciting direction involves understanding the behavior of strings at a \textit{finite non-asymptotic} Dirichlet boundary on a bulk Cauchy slice in $\operatorname{AdS}$ \cite{Zamolodchikov:2004ce,Smirnov:2016lqw,Cavaglia:2016oda,McGough:2016lol,Araujo-Regado:2022gvw} for T$\bar{\text{T}}$-deformed CFTs. The related question of placing a finite timelike boundary in an $\operatorname{AdS}$ black hole and in a $\operatorname{dS}$ static patch requires an understanding of the behavior of strings near totally absorbing walls with a Dirichlet boundary condition \cite{silverstein2023black,Batra:2024kjl}.\footnote{We thank Eva Silverstein and Aron Wall for a discussion of this point.} It would be very interesting if the work in this paper can be generalized to address this question.

\section*{Acknowledgements}
We are grateful to Aron Wall, Rifath Khan,  Shoaib Akhtar, Prahar Mitra, Ronak Soni, Alex Frenkel, Minjae Cho, Harold Erbin, and Daniel Thompson for insightful discussions and comments. AA is supported by The Royal Society and by  STFC Consolidated Grant No. ST/X000648/1. ZY is supported by an Internal Graduate Studentship of Trinity College, Cambridge and the AFOSR grant FA9550-19-1-0260 ``Tensor Networks and Holographic Spacetime''. The work in this paper was done while VS was a Postdoctoral Research Fellow at Stanford Institute for Theoretical Physics, Stanford University.
\paragraph{}
{\footnotesize {\bf Open Access Statement} - For the purpose of open access, the authors have applied a Creative Commons Attribution (CC BY) licence to any Author Accepted Manuscript version arising.}

\footnotesize{\textbf{Data access statement}: no new data were generated for this work.}

\bibliographystyle{JHEP}
\bibliography{main.bib}

\end{document}